\documentclass[amsmath,amsfonts,amssymb,twocolumn,nofootinbib]{revtex4-1}
\usepackage{graphicx}	
\usepackage{amssymb, amsmath}
\usepackage{subfigure}
\usepackage{hyperref}
\usepackage{color}
\usepackage[table]{xcolor}
\usepackage{multirow}
\usepackage{tikz}

\begin{document}

\title{Hamiltonian Monte Carlo for Hierarchical Models}

\author{Michael Betancourt}
\email{betanalpha@gmail.com}
\affiliation{Department of Statistical Science, University College London, London, UK}

\author{Mark Girolami}
\affiliation{Department of Statistical Science, University College London, London, UK}

\begin{abstract}
Hierarchical modeling provides a framework for modeling the complex interactions
typical of problems in applied statistics.  By capturing these relationships, however, 
hierarchical models also introduce distinctive pathologies that quickly limit the efficiency 
of most common methods of inference.  In this paper we explore the use of Hamiltonian 
Monte Carlo for hierarchical models and demonstrate how the algorithm can overcome 
those pathologies in practical applications.
\end{abstract}

\maketitle

Many of the most exciting problems in applied statistics involve intricate, typically
high-dimensional, models and, at least relative to the model complexity, sparse data.
With the data alone unable to identify the model, valid inference in these circumstances 
requires significant prior information.  Such information, however, is not limited to the 
choice of an explicit prior distribution: it can be encoded in the construction of the model itself.

Hierarchical models take this latter approach, associating parameters into exchangeable
groups that draw common prior information from shared parent groups. The interactions 
between the levels in the hierarchy allow the groups to learn from each other without having
to sacrifice their unique context, partially pooling the data together to improve inferences.
Unfortunately, the same structure that admits powerful modeling also induces formidable
pathologies that limit the performance of those inferences.

After reviewing hierarchical models and their pathologies, we'll discuss common 
implementations and show how those pathologies either make the algorithms impractical 
or limit their effectiveness to an unpleasantly small space of models. We then introduce 
Hamiltonian Monte Carlo and show how the novel properties of the algorithm can yield
much higher performance for general hierarchical models.  Finally we conclude with
examples which emulate the kind of models ubiquitous in contemporary applications.

\section{Hierarchical Models}

\begin{figure}
\setlength{\unitlength}{0.07in} 
\centering
\begin{picture}(40, 30)
\put(20, 5) { \circle{6} }
\put(20, 5) { \makebox(0, 0) { $\phi$ } }
\put(4, 15) { \circle{6} }
\put(4, 15) { \makebox(0, 0) { $\theta_{1}$ } }
\put(12, 15) { \circle{6} }
\put(12, 15) { \makebox(0, 0) { $\theta_{2}$ } }
\put(20, 15) { \circle{6} }
\put(20, 15) { \makebox(0, 0) { $\ldots$ } }
\put(28, 15) { \circle{6} }
\put(28, 15) { \makebox(0, 0) { $\theta_{n - 1}$ } }
\put(36, 15) { \circle{6} }
\put(36, 15) { \makebox(0, 0) { $\theta_{n}$ } }
\put(20, 8) { \vector(-4, 1){16} }
\put(20, 8) { \vector(-2, 1){8} }
\put(20, 8) { \vector(0, 1){4} }
\put(20, 8) { \vector(2, 1){8} }
\put(20, 8) { \vector(4, 1){16} }
\put(0.75, 21.75) { \tikz\draw[black,fill=black] circle (0.22in); }
\put(4, 25) { \makebox(0, 0) { \color{white} $\mathcal{D}_{1}$ } }
\put(8.75, 21.75) { \tikz\draw[black,fill=black] circle (0.22in); }
\put(12, 25) { \makebox(0, 0) { \color{white} $\mathcal{D}_{2}$ } }
\put(16.75, 21.75) { \tikz\draw[black,fill=black] circle (0.22in); }
\put(20, 25) { \makebox(0, 0) { \color{white} $\ldots$ } }
\put(24.75, 21.75) { \tikz\draw[black,fill=black] circle (0.22in); }
\put(28, 25) { \makebox(0, 0) { \color{white} $\mathcal{D}_{n - 1}$ } }
\put(32.75, 21.75) { \tikz\draw[black,fill=black] circle (0.22in); }
\put(36, 25) { \makebox(0, 0) { \color{white} $\mathcal{D}_{n}$ } }
\put(4, 18) { \vector(0, 1){4} }
\put(12, 18) { \vector(0, 1){4} }
\put(20, 18) { \vector(0, 1){4} }
\put(28, 18) { \vector(0, 1){4} }
\put(36, 18) { \vector(0, 1){4} }
\end{picture} 
\caption{
In hierarchical models ``local'' parameters, $\theta$, interact via a common dependency on 
``global'' parameters, $\phi$.  The interactions allow the measured data, $\mathcal{D}$, to 
inform all of the $\theta$ instead of just their immediate parent.  More general constructions 
repeat this structure, either over different sets of parameters or additional layers of hierarchy.
}
\label{fig:hierModel} 
\end{figure}
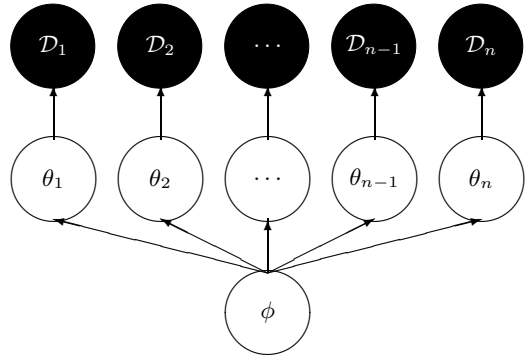

Hierarchical models~\cite{GelmanEtAl:2013} are defined by the organization of a model's parameters into 
exchangeable groups, and the resulting conditional independencies between those groups.%
\footnote{Not that all parameters have to be grouped into the same hierarchical structure.
Models with different hierarchical structures for different parameters are known as
\textit{multilevel models}.}
A one-level hierarchy with parameters $\left( \theta, \phi \right)$ and data $\mathcal{D}$, 
for example, factors as (Figure \ref{fig:hierModel})
\begin{equation} \label{oneLevel}
\pi \! \left( \theta, \phi | \mathcal{D} \right)
\propto \prod_{i = 1}^{n} \pi \! \left( \mathcal{D}_{i} | \theta_{i} \right) 
\pi \! \left( \theta_{i} | \phi \right) \pi \! \left( \phi \right).
\end{equation}
A common example is the one-way normal model,
\begin{align}
y_{i} &\sim \mathcal{N} \! \left(\theta_i, \sigma_{i}^{2} \right) \notag \\
\theta_i &\sim \mathcal{N} \! \left(\mu, \tau^2 \right), \mbox{ for } i=1,\dots,I, \label{normal}
\end{align}
or, in terms of the general notation of (\ref{oneLevel}), $\mathcal{D} = (y_{i}, \sigma_{i})$, 
$\phi=(\mu, \tau)$, and $\theta=(\theta_{i})$.  To ease exposition we refer to any elements of $\phi$ 
as global parameters, and any elements of $\theta$ as local parameters, even though such 
a dichotomy quickly falls apart when considering models with multiple layers.

Unfortunately for practitioners, but perhaps fortunately for pedagogy, the one-level model 
\eqref{oneLevel} exhibits all of the pathologies typical of hierarchical models.  Because the 
$n$ contributions at the bottom of the hierarchy all depend on the global parameters, a small 
change in $\phi$ induces large changes in the density.  Consequently, when the data are 
sparse the density of these models looks like a ``funnel'', with a region of high density but 
low volume below a region of low density and high volume.  The probability \textit{mass} of 
the two regions, however, is comparable and any successful sampling algorithm must be able 
to manage the dramatic variations in curvature in order to fully explore the posterior.

\begin{figure}
\centering
\includegraphics[width=2.9in]{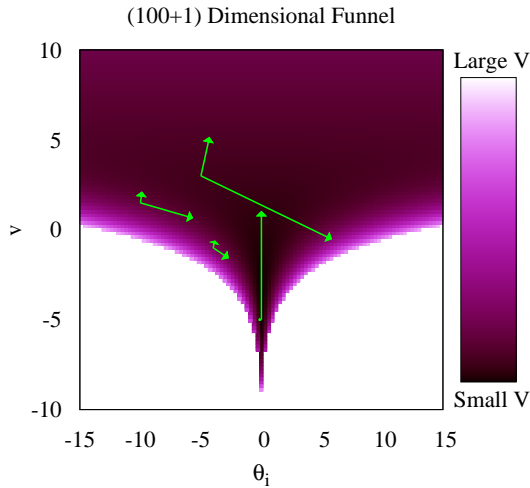}
\caption{
Typical of hierarchical models, the curvature of the funnel distribution varies strongly with the 
parameters, taxing most algorithms and limiting their ultimate performance.
Here the curvature is represented visually by the eigenvectors of 
$\sqrt{ \left| \partial^{2} \log p \!  \left( \theta_{1}, \ldots, \theta_{n}, v \right) / \partial \theta_{i} \partial \theta_{j} \right| }$ 
scaled by their respective eigenvalues, which encode the direction and magnitudes of the local
deviation from isotropy.
}
\label{fig:funnelCurvature} 
\end{figure}

For visual illustration, consider the funnel distribution~\cite{Neal:2003} resulting from a one-way 
normal model with no data, latent mean $\mu$ set to zero, and a log-normal prior on the 
variance $\tau^{2} = e^{v}$,%
\footnote{The exponential relationship between the latent $v$ and the variance $\tau^{2}$
may appear particularly extreme, but it arises naturally whenever one transforms from a
parameter constrained to be positive to an unconstrained parameter more appropriate
for sampling.}
\begin{equation*}
\pi \! \left( \theta_{1}, \ldots, \theta_{n}, v \right) 
\propto \prod_{i = 1}^{n} \mathcal{N} \! \left( x_{i} | 0, ( e^{-v/2} )^{2} \right) 
\mathcal{N} \! \left( v | 0, 3^{2} \right).
\end{equation*}
The hierarchical structure induces large correlations between $v$ and each of the $\theta_{i}$, 
with the correlation strongly varying with position (Figure \ref{fig:funnelCurvature}).  Note that the
position-dependence of the correlation ensures that no global correction, such as a rotation and
rescaling of the parameters, will simplify the distribution to admit an easier implementation.

\section{Common Implementations of Hierarchical Models}

Given the utility of hierarchical models, a variety of implementations have been developed
with varying degrees of success.  Deterministic algorithms, for example 
\cite{PinheiroEtAl:2000, RabeEtAl:2008, RueEtAl:2009}, can be quite powerful in a limited scope
of models.  Here we instead focus on the stochastic Markov Chain Monte Carlo algorithms, in
particular Metropolis and Gibbs samplers, which offer more breadth.

\subsection{Na\"{i}ve Implementations}

Although they are straightforward to implement for many hierarchical models, the performance of 
algorithms like Random Walk Metropolis and the Gibbs sampler \cite{RobertEtAl:1999} is limited
by their incoherent exploration.  More technically, these algorithms explore via transitions tuned to 
the conditional variances of the target distribution.  When the target is highly correlated, however, 
the conditional variances are much smaller than the marginal variances and many transitions are 
required to explore the entire distribution.  Consequently, the samplers devolve into random walks
which explore the target distribution extremely slowly.

\begin{figure*}
\centering
\subfigure[]{ \label{fig:gibbsFunnel} \includegraphics[width=6.5in]{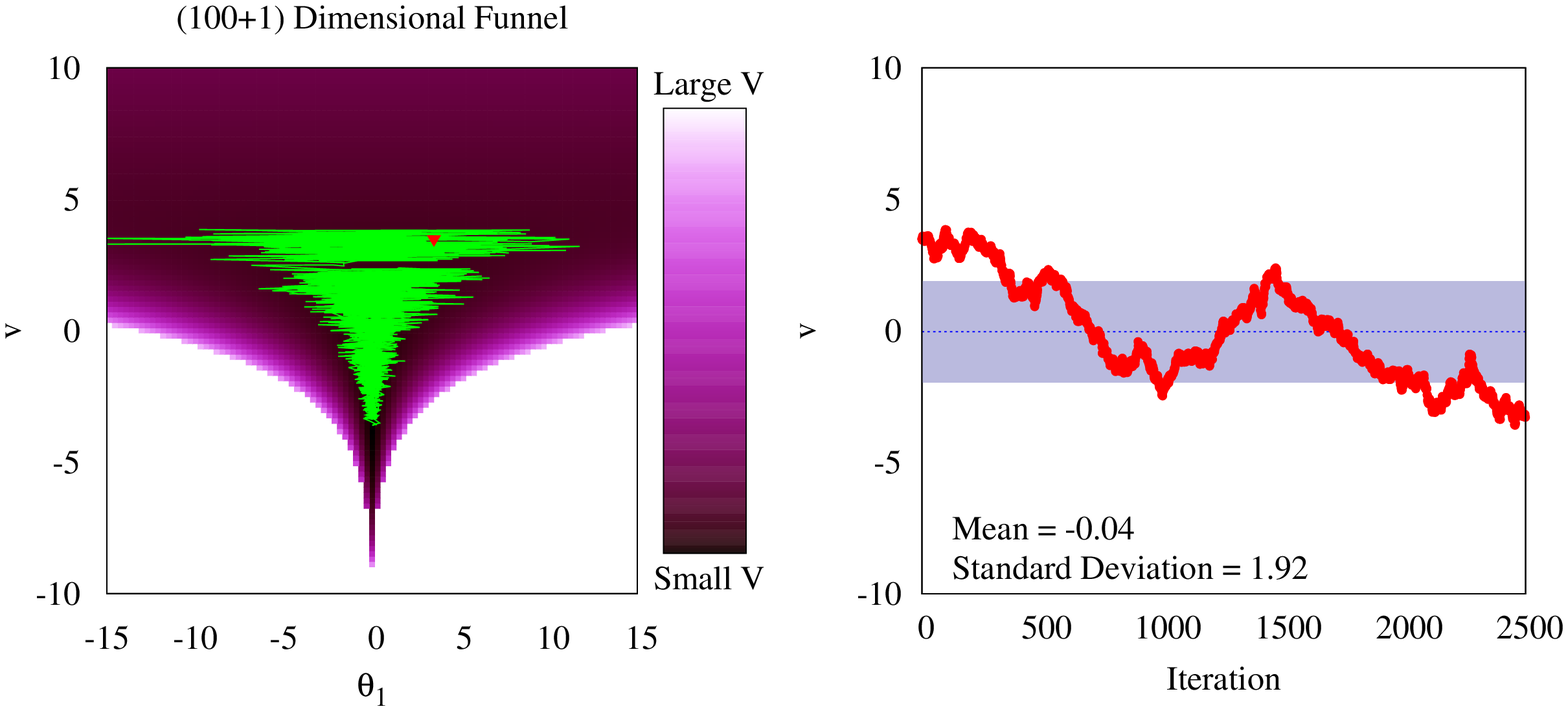}}
\subfigure[]{ \label{fig:metroFunnel} \includegraphics[width=6.5in]{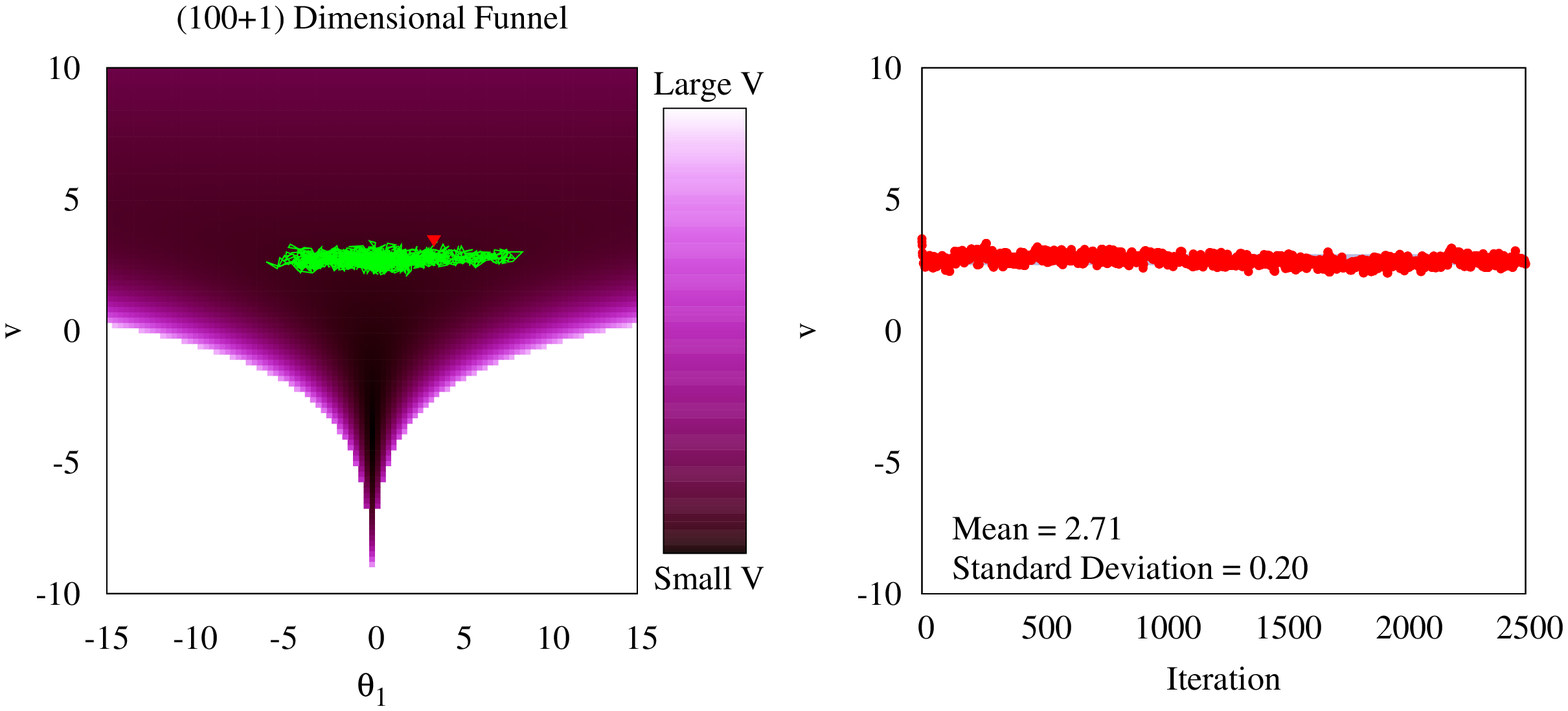}}
\caption{One of the biggest challenges with modern models
is not global correlations but rather local correlations
that are resistant to the corrections based on the global
covariance.  The funnel distribution, here with $N=100$,
features strongly varying local correlations but enough
symmetry that these correlations cancel globally, so no single
correction can compensate for the ineffective exploration
of (a) the Gibbs sampler and (b) Random Walk Metropolis.  After
2500 iterations neither chain has explored the marginal distribution
of $v$, $\pi \! \left( v \right) = \mathcal{N} \! \left(v | 0, 3^{2} \right)$.
Note that the Gibbs sampler utilizes a Metropolis-within-Gibbs
scheme as the conditional $\pi \! \big( v | \vec{\theta} \big)$ does
not have a closed-form.
}
\label{fig:funnel}
\end{figure*}

As we saw above, hierarchical models are highly correlated by construction.  As more groups 
and more levels are added to the hierarchy, the correlations worsen and na\"{i}ve MCMC implementations
quickly become impractical (Figure \ref{fig:funnel}).

\subsection{Efficient Implementations}

A common means of improving Random Walk Metropolis and the Gibbs sampler is to correct for
global correlations, bringing the conditional variances closer to the marginal variances and reducing 
the undesired random walk behavior.  The correlations in hierarchical models, however, are not global 
but rather local and efficient implementations require more sophistication.  To reduce the correlations 
between successive layers and improve performance we have to take advantage of the hierarchical structure 
explicitly.  Note that, because this structure is defined in terms of conditional independencies, these strategies 
tend to be more natural, not to mention more successful, for Gibbs samplers.

One approach is to separate each layer with auxiliary variables 
\cite{LiuEtAl:1998, LiuEtAl:1999, GelmanEtAl:2008}, for example the one-way normal model 
\eqref{normal} would become
\begin{equation*}
\theta_i = \mu + \xi\eta_i,\ \eta_i\sim \mathcal{N} \! \left(0,\sigma^2_{\eta} \right),
\end{equation*}
with $\tau=|\xi|\sigma_{\eta}$.  Conditioned on $\eta$, the layers become independent and the Gibbs
sampler can efficiently explore the target distribution.  On the other hand, the multiplicative
dependence of the auxiliary variable actually introduces strong correlations into the joint distribution
that diminishes the performance of Random Walk Metropolis.
 
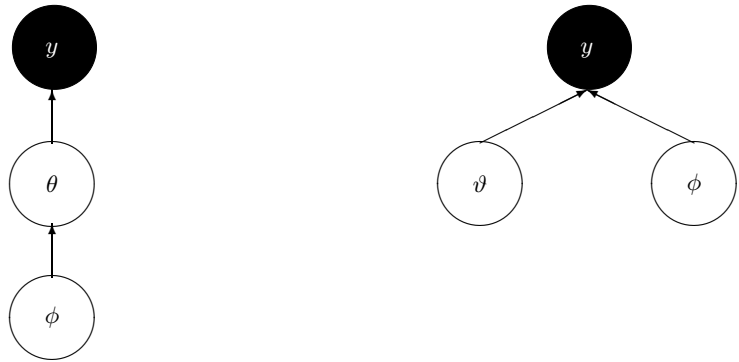
\begin{figure*}
\setlength{\unitlength}{0.07in} 
\centering
\begin{picture}(80, 30)
\put(20, 5) { \circle{6} }
\put(20, 5) { \makebox(0, 0) { $\phi$ } }
\put(20, 15) { \circle{6} }
\put(20, 15) { \makebox(0, 0) { $\theta$ } }
\put(20, 8) { \vector(0, 1){4} }
\put(17, 22) { \tikz\draw[black,fill=black] circle (0.22in); }
\put(20, 25) { \makebox(0, 0) { \color{white} $y$ } }
\put(20, 18) { \vector(0, 1){4} }
\put(52, 15) { \circle{6} }
\put(52, 15) { \makebox(0, 0) { $\vartheta$ } }
\put(68, 15) { \circle{6} }
\put(68, 15) { \makebox(0, 0) { $\phi$ } }
\put(52, 18) { \vector(2, 1){8} }
\put(68, 18) { \vector(-2, 1){8} }
\put(57, 22) { \tikz\draw[black,fill=black] circle (0.22in); }
\put(60, 25) { \makebox(0, 0) { \color{white} $y$ } }
\end{picture} 
\caption{
In one-level hierarchical models with global parameters, $\phi$, local parameters, $\theta$,
and measured data $y$, correlations between parameters can be mediated by different
parameterizations of the model.  Non-centered parameterizations exchange a direct
dependence between $\phi$ and $\theta$ for a dependence between $\phi$ and $y$;
the reparameterized $\vartheta$ and $\phi$ become independent conditioned on the data.
When the data are weak these non-centered parameterizations yield simpler posterior
geometries.
}
\label{fig:reparam} 
\end{figure*}

In addition to adding new parameters, the dependence between layers can also be broken by
reparameterizing existing parameters.  Non-centered parameterizations, for example 
\cite{PapaspiliopoulosEtAl:2007}, factor certain dependencies into deterministic transformations
between the layers, leaving the actively sampled variables uncorrelated (Figure \ref{fig:reparam}).  
In the one-way normal model \eqref{normal} we would apply both location and scale 
reparameterizations yielding
\begin{align*}
y_{i} &\sim \mathcal{N} \! \left( \vartheta_{i} \tau + \mu, \sigma_{i}^{2} \right) \\
\vartheta_i &\sim \mathcal{N} \! \left(0, 1\right),
\end{align*}
effectively shifting the correlations from the latent parameters to the data.  Provided that the the 
data are not particularly constraining, i.e. the $\sigma_{i}^{2}$ are large, the resulting joint 
distribution is almost isotropic and the performance of both Random Walk Metropolis and the 
Gibbs sampler improves.

Unfortunately, the ultimate utility of these efficient implementations is limited to the relatively small
class of models where analytic results can be applied.  Parameter expansion, for example, requires
that the expanded conditional distributions can be found in closed form (although, to be fair, that is also
a requirement for any Gibbs sampler in the first place) while non-centered parameterizations are
applicable mostly to models where the dependence between layers is given by a generalized linear
model.  To enable efficient inference without constraining the model we need to consider more
sophisticated Markov Chain Monte Carlo techniques.

\section{Hamiltonian Monte Carlo for Hierarchical Models}

Hamiltonian Monte Carlo \cite{DuaneEtAl:1987, Neal:2011, BetancourtEtAl:2011} utilizes techniques from 
differential geometry to generate transitions spanning the full marginal variance, eliminating the random walk 
behavior endemic to Random Walk Metropolis and the Gibbs samplers.

The algorithm introduces auxiliary \textit{momentum} variables, $p$, to the parameters of the target distribution, 
$q$, with the joint density
\begin{equation*}
\pi \! \left( p, q \right) = \pi \! \left( p | q \right) \pi \! \left( q \right).
\end{equation*}
After specifying the conditional density of the momenta, the joint density defines a Hamiltonian,
\begin{align*}
H \! \left( p, q \right) 
&= - \log \pi \! \left(p, q \right) \\
&= - \log \pi \! \left(p | q \right) - \log \pi \! \left(q \right) \\
&= T \! \left(p | q \right) + V \! \left(q \right),
\end{align*}
with the \textit{kinetic energy},
\begin{equation*}
T \! \left(p | q \right) \equiv - \log \pi \! \left(p | q \right),
\end{equation*}
and the \textit{potential energy},
\begin{equation*}
V \! \left(q \right) \equiv - \log \pi \! \left(q \right).
\end{equation*}
This Hamiltonian function generates a transition by first sampling the auxiliary momenta,
\begin{equation*}
p \sim \pi \! \left( p | q \right),
\end{equation*}
and then evolving the joint system via Hamilton's equations,
\begin{align*}
\frac{dq}{dt} &= +\frac{ \partial H }{ \partial p } = + \frac{ \partial T }{ \partial p } \\
\frac{dp}{dt} &= -\frac{ \partial H }{ \partial q } = - \frac{ \partial T }{ \partial q } - \frac{ \partial V }{ \partial q }.
\end{align*}
The gradients guide the transitions through regions of high probability and admit the efficient 
exploration of the entire target distribution.  For how long to evolve the system depends on 
the shape of the target distribution, and the optimal value may vary with position~\cite{Betancourt:2013b}.  
Dynamically determining the optimal integration time is highly non-trivial as na\"{i}ve implementations break the 
detailed balance of the transitions; the No-U-Turn sampler preserves detailed balance by integrating
not just forward in time but also backwards~\cite{HoffmanEtAl:2011}.

Because the trajectories are able to span the entire marginal variances, the efficient exploration of
Hamiltonian Monte Carlo transitions persists as the target distribution becomes correlated, and even 
if those correlations are largely local, as is typical of hierarchical models.  As we will see, however, 
depending on the choice of the momenta distribution, $\pi \! \left(p | q \right)$, hierarchical models 
can pose their own challenges for Hamiltonian Monte Carlo.

\begin{figure*}
\centering
\subfigure[]{\includegraphics[width=2.25in]{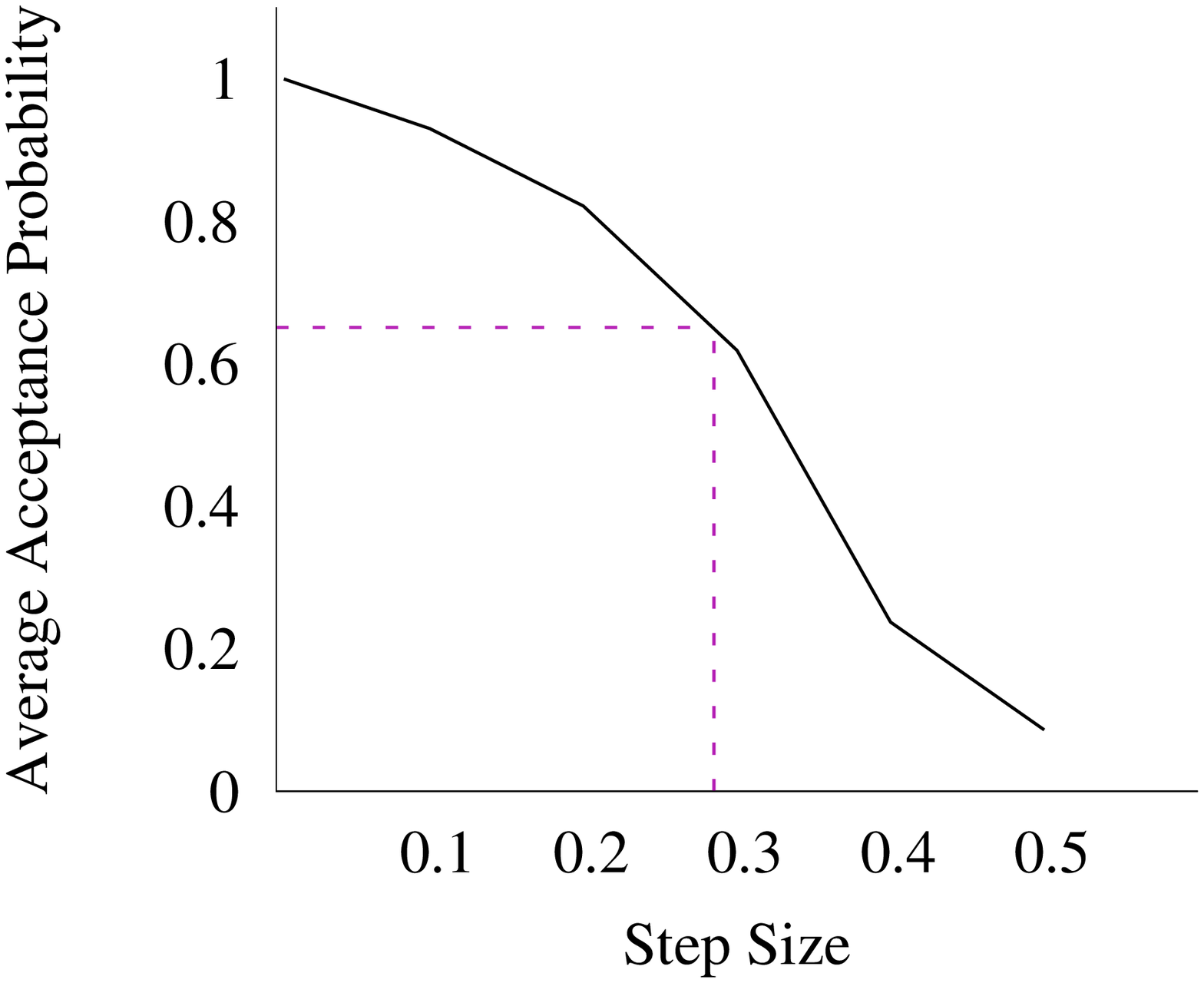}}
\subfigure[]{\includegraphics[width=2.25in]{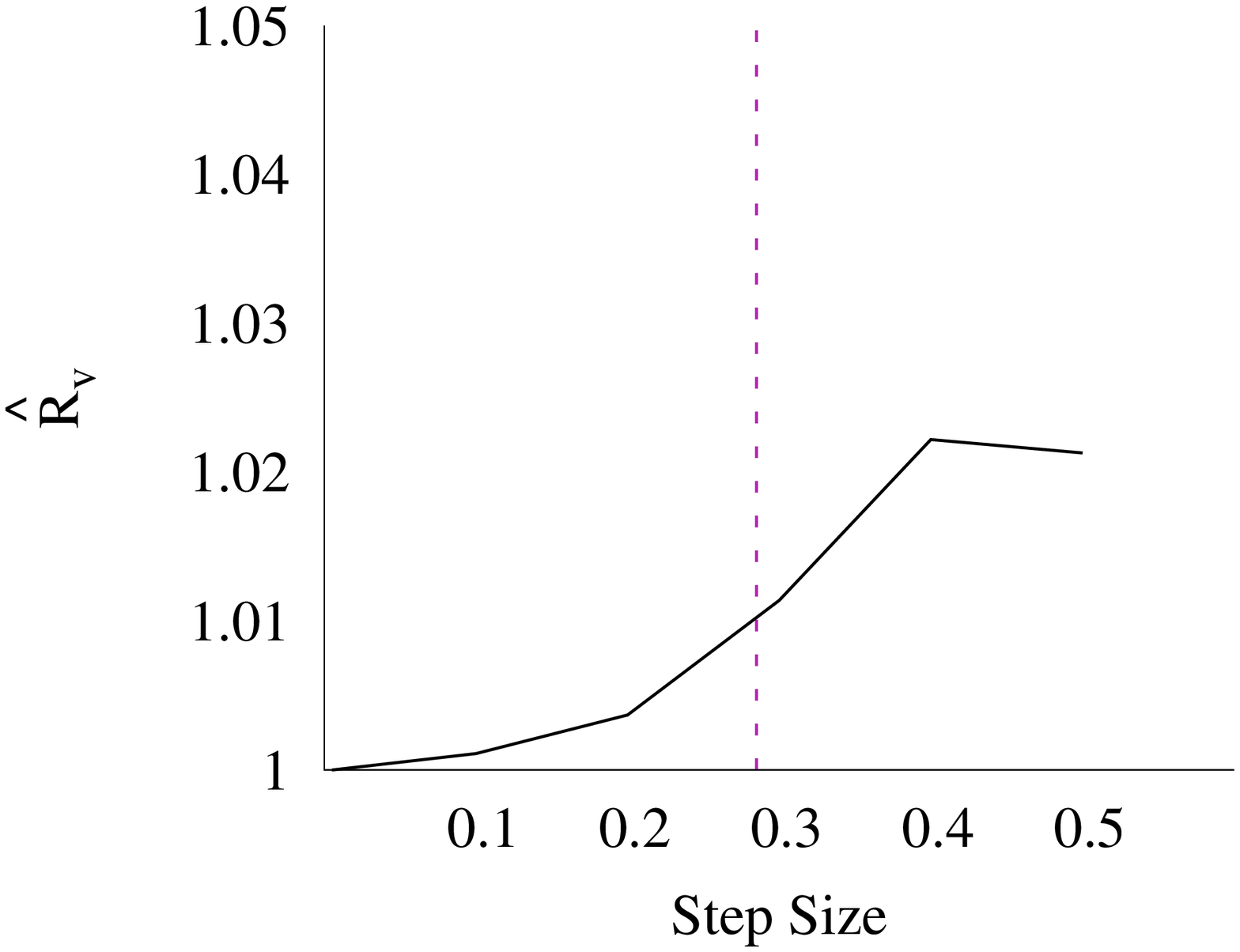}}
\subfigure[]{\includegraphics[width=2.25in]{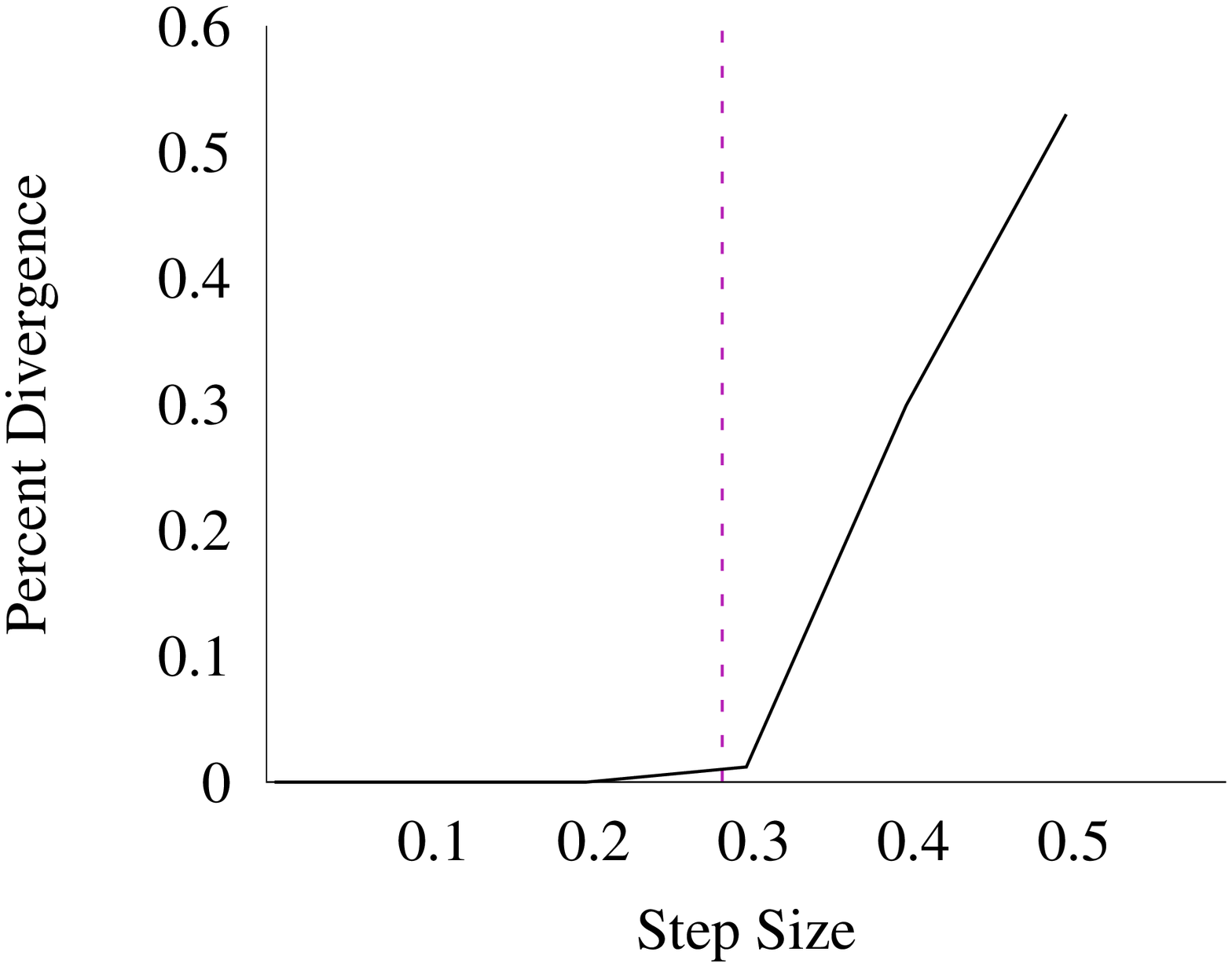}}
\caption{Careful consideration of any adaptation procedure is crucial for valid inference in hierarchical models.
As the step size of the numerical integrator is decreased (a) the average acceptance probability increases
from the canonically optimal value of 0.651 but (b) the sampler output converges to a consistent distribution. Indeed,
(c) at the canonically optimal value of the average acceptance probability the integrator begins to diverge.  
Here consistency is measured with a modified potential scale reduction statistic~\cite{GelmanEtAl:1992, StanMan:2013} 
for the latent parameter $v$ in a (50 + 1)-dimensional funnel.
}
\label{fig:funnelScan}
\end{figure*}

\subsection{Euclidean Hamiltonian Monte Carlo}

The simplest choice of the momenta distribution, and the one almost exclusively seen in contemporary applications, 
is a Gaussian independent of $q$,
\begin{equation*}
\pi \! \left( p | q \right) = \mathcal{N} \! \left( p | 0, \Sigma \right),
\end{equation*}
resulting in a quadratic kinetic energy,
\begin{equation*}
T \! \left(p, q \right) = \frac{1}{2} p^{T} \Sigma^{-1} p.
\end{equation*}
Because the subsequent Hamiltonian also generates dynamics on a Euclidean manifold, we refer to the resulting 
algorithm at Euclidean Hamiltonian Monte Carlo.  Note that the metric, $\Sigma$, effectively induces 
a global rotation and rescaling of the target distribution, although it is often taken to be the identity
in practice. 

Despite its history of success in difficult applications, Euclidean Hamiltonian Monte Carlo does have two 
weaknesses that are accentuated in hierarchical models: the introduction of a characteristic length scale and 
limited variations in density.

\subsubsection{Characteristic Length Scale} \label{sec:lengthScale}

In practice Hamilton's equations are sufficiently complex to render analytic solutions infeasible; 
instead the equations must be integrated numerically.  Although symplectic integrators provide 
efficient and accurate numerical solutions~\cite{HairerEtAl:2006}, they introduce a characteristic 
length scale via the time discretization, or step size, $\epsilon$.

Typically the step size is tuned to achieve an optimal acceptance probability~\cite{BeskosEtAl:2013}, 
but such optimality criteria ignore the potential instability of the integrator.  In order to prevent the 
numerical solution from diverging before it can explore the entire distribution, the step size must be 
tuned to match the curvature.  Formally, a stable solution requires~\cite{HairerEtAl:2006}
\begin{equation*}
\epsilon \sqrt{ \lambda_{i}  } < 2
\end{equation*}
for each eigenvalue, $\lambda_{i}$, of the matrix%
\footnote{Note that the ultimate computational efficiency of Euclidean Hamiltonian Monte Carlo scales with the condition 
number of $M$ averaged over the target distribution.  A well chosen metric reduces the condition number, explaining why
a global decorrelation that helps with Random Walk Metropolis and Gibbs sampling is also favorable for Euclidean Hamiltonian 
Monte Carlo.}
\begin{equation*}
M_{ij} =  \left( \Sigma^{-1} \right)_{ik} \frac{ \partial^{2} V }{ \partial q_{k} \partial q_{j} }.
\end{equation*}

Moreover, algorithms that adapt the step size to achieve an optimal acceptance probability require a 
relatively precise and accurate estimate of the global acceptance probability.  When the chain has 
high autocorrelation or overlooks regions of high curvature because of a divergent integrator, 
however, such estimates are almost impossible to achieve.  Consequently, adaptive algorithms 
can adapt too aggressively to the local neighborhood where the chain was seeded, potentially 
biasing resulting inferences.

Given the ubiquity of spatially-varying curvature, these pathologies are particularly common to 
hierarchical models. In order to use adaptive algorithms we recommend relaxing the adaptation 
criteria to ensure that the Markov chain hasn't been biased by overly assertive adaptation.  A 
particularly robust strategy is to compare inferences, especially for the latent parameters, as the 
adaptation criteria is gradually weakened, selecting a step size only once the inferences have 
stabilized and divergent transitions are rare (Figure \ref{fig:funnelScan}).  At the very least an auxiliary 
chain should always be run at a smaller step size to ensure consistent inferences.

\subsubsection{Limited Density Variations}

\begin{figure}
\centering
\includegraphics[width=3in]{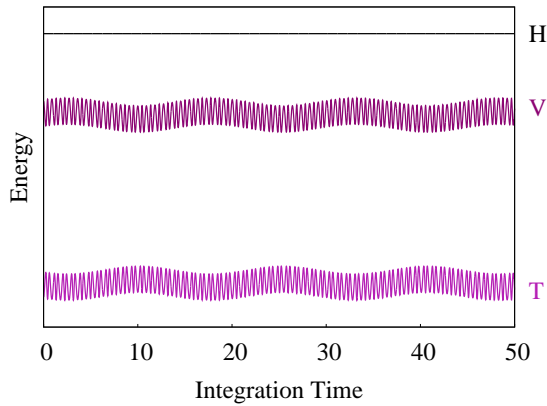}
\caption{Because the Hamiltonian, $H$, is conserved during each trajectory, the variation
in the potential energy, $V$, is limited to the variation in the kinetic energy, $T$, which itself
is limited to only $d / 2$. }
\label{fig:ehmcTrace}
\end{figure}

\begin{figure*}
\centering
\includegraphics[width=6.5in]{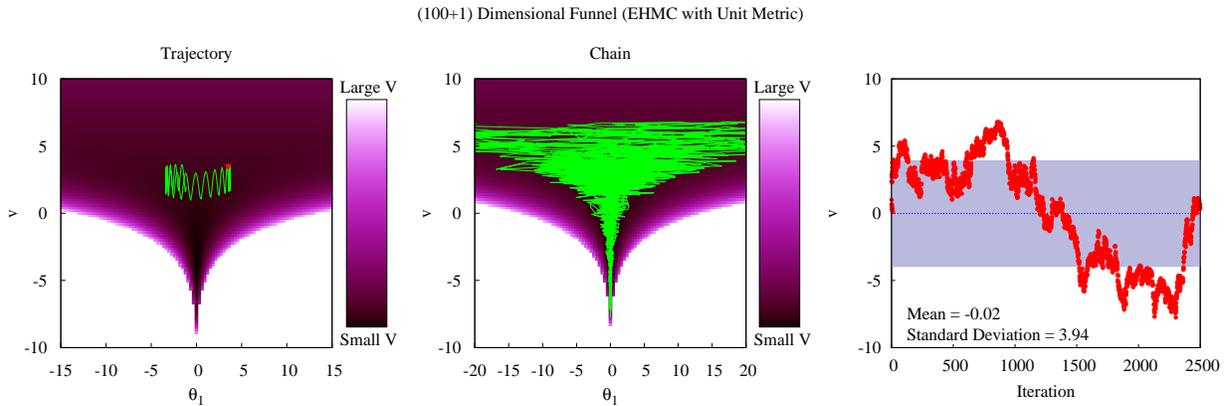}
\caption{Limited to moderate potential energy variations, the trajectories of Euclidean HMC, here with a unit metric
$\Sigma = \mathbb{I}$, reduce to random walk behavior in hierarchical models.  The resulting Markov chain explores 
more efficiently than Gibbs and Random Walk Metropolis (Figure \ref{fig:funnel}), but not efficiently enough to make these models particularly
practical.}
\label{fig:ehmcFunnel}
\end{figure*}

A more subtle, but no less insidious, vulnerability of Euclidean HMC concerns density variations with a transition.  
In the evolution of the system, the Hamiltonian function,
\begin{equation*}
H \! \left(p, q \right) = T \! \left(p | q \right) + V \! \left(q \right),
\end{equation*}
is constant, meaning that any variation in the potential energy must be compensated for by an opposite variation 
in the kinetic energy.  In Euclidean Hamiltonian Monte Carlo, however, the kinetic energy is a $\chi^{2}$ variate
which, in expectation, varies by only half the dimensionality, $d$, of the target distribution.  Consequently the 
Hamiltonian transitions are limited to
\begin{equation*}
\Delta V = \Delta T \sim \frac{d}{2},
\end{equation*}
restraining the density variation within a single transition.  Unfortunately the correlations inherent to hierarchical models 
also induce huge density variations, and for any but the smallest hierarchical model this restriction prevents
the transitions from spanning the full marginal variation.  Eventually random walk behavior creeps back in
and the efficiency of the algorithm plummets (Figure \ref{fig:ehmcTrace}, \ref{fig:ehmcFunnel}).

\begin{figure}
\centering
\includegraphics[width=3in]{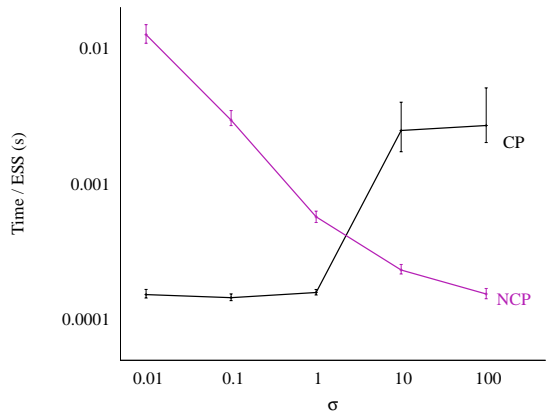}
\caption{Depending on the common variance, $\sigma^{2}$, from which the data were generated, the performance
of a 10-dimensional one-way normal model \eqref{normal} varies drastically between centered (CP) and non-centered (NCP)
parameterizations of the latent parameters, $\theta_{i}$.  As the variance increases and the data become effectively more
sparse, the non-centered parameterization yields the most efficient inference and the disparity in performance increases
with the dimensionality of the model.  The bands denote the quartiles over an ensemble of 50 runs, with each run using 
Stan~\cite{Stan:2013} configured with a diagonal metric and the No-U-Turn sampler.  Both the metric and the step size were 
adapted during warmup, and care was taken to ensure consistent estimates (Figure \ref{fig:performanceTuning}).}
\label{fig:paramPerformance}
\end{figure}

Because they remove explicit hierarchical correlations, non-centered parameterizations can also reduce the density variations 
of hierarchical models and drastically increase the performance of Euclidean Hamiltonian Monte Carlo 
(Figure \ref{fig:paramPerformance}).  Note that as in the case of Random Walk Metropolis and Gibbs sampling, the efficacy 
of the parametrization depends on the relative strength of the data, although when the nominal centered parameterization is 
best there is often enough data that the partial pooling of hierarchical models isn't needed in the first place.

\begin{figure}
\centering
\includegraphics[width=2.5in]{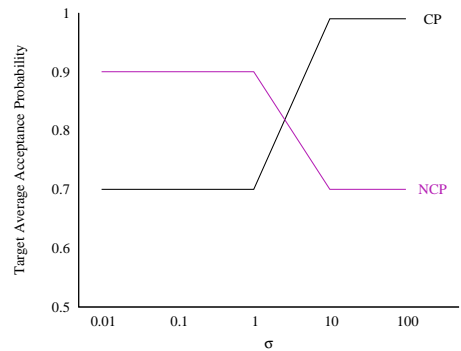}
	\caption{As noted in Section \ref{sec:lengthScale}, care must be taken when using adaptive implementations of Euclidean
	Hamiltonian Monte Carlo with hierarchical models.  For the results in Figure \ref{fig:paramPerformance}, the optimal average 
	acceptance probability was relaxed until the estimates of $\tau$ stabilized, as measured by the potential scale reduction
	factor, and divergent transitions did not appear.}
\label{fig:performanceTuning}
\end{figure}

\subsection{Riemannian Hamiltonian Monte Carlo}

Euclidean Hamiltonian Monte Carlo is readily generalized by allowing the covariance to vary with position 
\begin{equation*}
\pi \! \left( p | q \right) = \mathcal{N} \! \left( p | 0, \Sigma \! \left(q\right) \right),
\end{equation*}
giving,
\begin{equation*}
T \! \left(p, q \right) = \frac{1}{2} p^{T} \Sigma^{-1} \left( q \right) p - \frac{1}{2} \log \left| \Sigma \! \left( q \right) \right|.
\end{equation*}
With the Hamiltonian now generating dynamics on a  Riemannian manifold with metric $\Sigma$,
we follow the convention established above and denote the resulting algorithm as Riemannian Hamiltonian
Monte Carlo~\cite{GirolamiEtAl:2011}.

The dynamic metric effectively induces local corrections to the target distribution, and if the metric is well chosen 
then those corrections can compensate for position-dependent correlations, not only reducing the computational burden 
of the Hamiltonian evolution but also relieving the sensitivity to the integrator step size.

Note also the appearance of the log determinant term, $\frac{1}{2} \log \left| \Sigma \! \left( q \right) \right|$.  
Nominally in place to provide the appropriate normalization for the momenta distribution, this term 
provides a powerful feature to Riemannian Hamiltonian Monte Carlo, serving as a reservoir that 
absorbs and then releases energy along the evolution and potentially allowing much larger variations 
in the potential energy.

\begin{figure}
\centering
\includegraphics[width=2.6in]{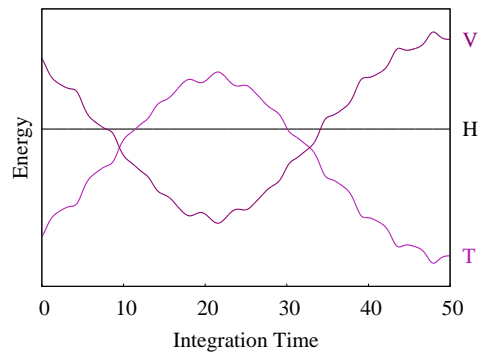}
\caption{Although the variation in the potential energy, $V$, is still limited by the variation in the kinetic energy, 
$T$, the introduction of the log determinant term in Riemannian Hamiltonian Monte Carlo allows the kinetic
energy sufficiently large variation that the potential is essentially unconstrained in practice. }
\label{fig:rhmcTrace}
\end{figure}

\begin{figure*}
\centering
\includegraphics[width=6.5in]{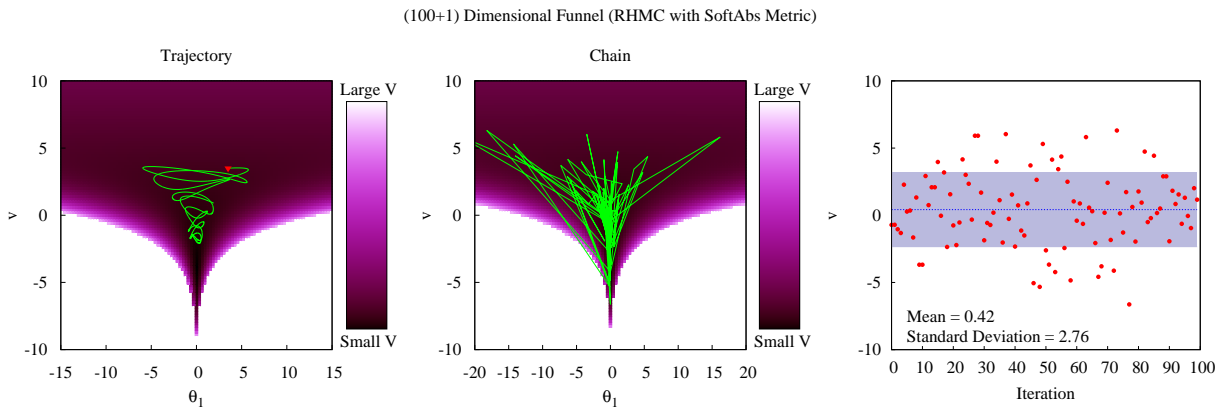}
\caption{Without being limited to small variations in the potential energy, Riemannian Hamiltonian Monte Carlo
with the SoftAbs metric admits transitions that expire the entirety of the funnel distribution, resulting in nearly
independent transitions, and drastically smaller autocorrelations (compare with Figure \ref{fig:ehmcFunnel},
noting the different number of iterations).}
\label{fig:rhmcFunnel}
\end{figure*}

Of course, the utility of Riemannian Hamiltonian Monte Carlo is dependent on the choice of the metric 
$\Sigma \! \left( q \right)$.  To optimize the position-dependent corrections we want a metric that
leaves the target distribution locally isotropic, motivating a metric resembling the Hessian of the target 
distribution.  Unfortunately the Hessian isn't sufficiently well-behaved to serve as a metric itself; in 
general, it is not even guaranteed to be positive-definite.  The Hessian can be manipulated into a 
well-behaved form, however, by applying the SoftAbs transformation,%
\footnote{Another approach to regularizing the Hessian is with the Fisher-Rao metric from information 
geometry~\cite{GirolamiEtAl:2011}, but this metric is able to regularize only by integrating out exactly the correlations 
needed for effective corrections, especially in hierarchical models~\cite{Betancourt:2013a}.}
and the resulting SoftAbs metric~\cite{Betancourt:2013a} admits a generic but efficient Riemannian Hamiltonian Monte 
Carlo implementation (Figure \ref{fig:rhmcTrace}, \ref{fig:rhmcFunnel}).

\pagebreak

\section{Example}

To see the advantage of Hamiltonian Monte Carlo over algorithms that explore with a random walk,
consider the one-way normal model \eqref{normal} with 800 latent $\theta_{i}$ and a constant measurement
error, $\sigma_{i} = \sigma$ across all nodes.  The latent parameters are taken to be $\mu = 8$ and $\tau = 3$,
with the $\theta_{i}$ and $y_{i}$ randomly sampled in turn with $\sigma = 10$.  To this generative likelihood
we add weakly-informative priors,
\begin{align*}
\pi \! \left( \mu \right) &= \mathcal{N} \! \left( 0, 5^{2} \right) \\
\pi \! \left( \tau \right) &= \text{Half-Cauchy} \! \left( 0, 2.5 \right).
\end{align*}
All sampling is done on a fully unconstrained space, in particular the latent $\tau$ is transformed
into $\lambda = \log \tau$.

\begin{figure}
\centering
\includegraphics[width=3in]{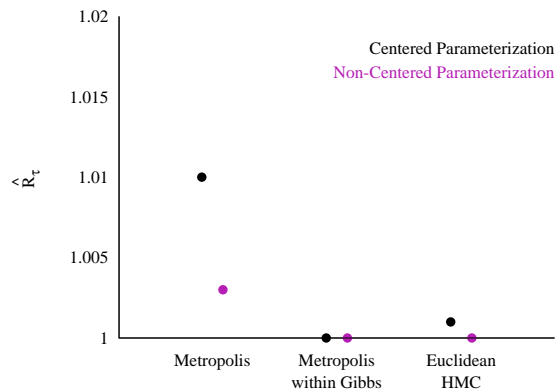}
\caption{In order to ensure valid comparisons, each sampling algorithm was optimized but only so long as the
resulting estimates were consistent with each other.}
\label{fig:exampleResults}
\end{figure}

Noting the results of Figure \ref{fig:paramPerformance}, the nominal centered parameterization,
\begin{align*}
y_{i} &\sim \mathcal{N} \! \left(\theta_i, \sigma_{i}^{2} \right) \\
\theta_i &\sim \mathcal{N} \! \left(\mu, \tau^2 \right), \mbox{ for } i=1,\dots,800,
\end{align*}
should yield inferior performance to the non-centered parameterization,
\begin{align}
y_{i} &\sim \mathcal{N} \! \left(\tau \vartheta_i + \mu, \sigma_{i}^{2} \right) \\
\vartheta_i &\sim \mathcal{N} \! \left(0, 1 \right), \mbox{ for } i=1,\dots,800;
\end{align}
in order to not overestimate the success of Hamiltonian Monte Carlo we include both.  For both parameterizations, 
we fit Random Walk Metropolis, Metropolis-within-Gibbs,%
\footnote{Because of the non-conjugate prior distributions the conditions for this model are not
analytic and we must resort to a Metropolis-within-Gibbs scheme as is common in practical applications.}
and Euclidean Hamiltonian Monte Carlo with a diagonal metric%
\footnote{Hamiltonian Monte Carlo is able to obtain more accurate estimates of the marginal variances 
than Random Walk Metropolis and Metropolis-within-Gibbs and, in a friendly gesture, the
variance estimates from Hamiltonian Monte Carlo were used to scale the transitions in the competing algorithms.}
to the generated data.%
\footnote{Riemannian Hamiltonian Monte Carlo was not considered here as its implementation in Stan
is still under development.}
The step size parameter in each case was tuned to be as large as possible whilst still
yielding consistent estimates with a baseline sample generated from running Euclidean
Hamiltonian Monte Carlo with a large target acceptance probability (Figure \ref{fig:exampleResults}).

Although the pathologies of the centered parameterization penalize all three algorithms, Euclidean Hamiltonian Monte
Carlo proves to be at least an order-of-magnitude more efficient than both Random Walk Metropolis and the Gibbs sampler.  
The real power of Hamiltonian Monte Carlo, however, is revealed when those penalties are removed in the non-centered 
parameterization (Table \ref{tab:exampleResults}).  

Most importantly, the advantage of Hamiltonian Monte Carlo scales with increasing dimensionality of the model.
In the most complex models that populate the cutting edge of applied statistics, Hamiltonian Monte Carlo is not just the
most convenient solution, it is often the only practical solution.

\begin{table*}
	\centering
	\renewcommand{\arraystretch}{1.5}
	\begin{tabular}{c@{}c@{}c@{}c@{}c@{}c}
	\rowcolor[gray]{0.95}  \hspace{2mm} \textbf{Algorithm} \hspace{2mm} 
	& \hspace{2mm} \textbf{Parameterization} \hspace{2mm} &  \hspace{2mm} \textbf{Step Size}  \hspace{2mm}
	&  \hspace{2mm} \textbf{Average}  \hspace{2mm} &  \hspace{2mm} \textbf{Time}  \hspace{2mm} 
	&  \hspace{2mm}\textbf{Time/ESS}  \hspace{2mm} \\
	\rowcolor[gray]{0.95} & & & \textbf{Acceptance} & \textbf{(s)} & \textbf{(s)} \\	
	\rowcolor[gray]{0.95} & & & \textbf{Probability} & & \\	
	Metropolis & Centered & $5.00 \cdot 10^{-3}$ & 0.822 & $4.51 \cdot 10^{4}$ & 1220 \\
	Gibbs & Centered & 1.50 & 0.446 & $9.54 \cdot 10^{4}$ & 297 \\
	EHMC & Centered & $1.91 \cdot 10^{-2}$ & 0.987 & $1.00 \cdot 10^{4}$ & \textbf{16.2} \\
	& & & & & \\
	Metropolis & Non-Centered & 0.0500 & 0.461 & 398 & 1.44 \\
	Gibbs & Non-Centered & 2.00 & 0.496 & 817 & 1.95 \\
	EHMC & Non-Centered & 0.164 & 0.763 & 154 & $\mathbf{2.94 \cdot 10^{-2} }$ \\
	\end{tabular}
	\label{tab:exampleResults}
	\caption{Euclidean Hamiltonian Monte Carlo significantly outperforms both Random Walk Metropolis 
	and the Metropolis-within-Gibbs sampler for both parameterizations of the one-way normal model.
	The difference is particularly striking for the more efficient non-centered parameterization that would
	be used in practice.}
\end{table*}

\section{Conclusion}

By utilizing the local curvature of the target distribution, Hamiltonian Monte Carlo provides the efficient
exploration necessary for learning from the complex hierarchical models of interest in applied problems.  
Whether using Euclidean Hamiltonian Monte Carlo with careful parameterizations or Riemannian Hamiltonian 
Monte Carlo with the SoftAbs metric, these algorithms admit inference whose performance scales not just with 
the size of the hierarchy but also with the complexity of local distributions, even those that may not be amenable 
to analytic manipulation.

The immediate drawback of Hamiltonian Monte Carlo is increased difficulty of implementation: not only
does the algorithm require non-trivial tasks such as the the integration of Hamilton's equations, the user must 
also specify the derivatives of the target distribution.  The inference engine Stan~\cite{Stan:2013} removes these burdens, 
providing not only a powerful probabilistic programming language for specifying the target distribution
and a high performance implementation of Hamiltonian evolution but also state-of-the-art automatic differentiation techniques
to compute the derivatives without any user input.  Through Stan, users can build, test, and run hierarchical models
without having to compromise for computational constraints.

\section{Acknowledgements}

We are indebted to Simon Byrne, Bob Carpenter, Michael Epstein, Andrew Gelman, Yair Ghitza, Daniel Lee, 
Peter Li, Sam Livingstone, and Anne-Marie Lyne for many fruitful discussions as well as invaluable comments 
on the text.  Michael Betancourt is supported under EPSRC grant EP/J016934/1 and Mark Girolami is an 
EPSRC Research Fellow under grant EP/J016934/1.

\appendix

\section{Stan Models}

One advantage of Stan is the ability to easily share and reproduce models.  Here we include all
models used in the tests above; results can be reproduced using the development branch
\url{https://github.com/stan-dev/stan/tree/f5df6e139df606c03bf02f4ff99d2341c15f73cd}
which will be incorporated into Stan v.2.1.0.

\subsection*{Funnel}

\subsubsection*{Model}

\footnotesize
\begin{verbatim}
transformed data {
  int<lower=0> J;
  J <- 25;
}
parameters {
  real theta[J];
  real v;
}
model {
  v ~ normal(0, 3);
  theta ~ normal(0, exp(v/2));
}
\end{verbatim}
\vspace{50mm}

\pagebreak

\subsection*{Generate One-Way Normal Pseudo-data}

\subsubsection*{Model}

\begin{verbatim}
transformed data {
  real mu;
  real<lower=0> tau;

  real alpha;
  int N;

  mu <- 8;
  tau <- 3;

  alpha <- 10;
  N <- 800;

}

parameters {
  real x;
}

model {
  x ~ normal(0, 1);
}

generated quantities {
  real mu_print;
  real tau_print;

  vector[N] theta;
  vector[N] sigma;
  vector[N] y;

  mu_print <- mu;
  tau_print <- tau;

  for (i in 1:N) {
    theta[i] <- normal_rng(mu, tau);
    sigma[i] <- alpha;
    y[i] <- normal_rng(theta[i], sigma[i]);
  }

}
\end{verbatim}
\subsubsection*{Configuration}
\begin{verbatim}
./generate_big_psuedodata sample num_warmup=0
num_samples=1 random seed=48383823 
output file=samples.big.csv 
\end{verbatim}
\vspace{50mm}

\pagebreak

\subsection*{One-Way Normal (Centered)}
\subsubsection*{Model}
\vspace{-1mm}
\begin{verbatim}
data {
  int<lower=0> J;
  real y[J];
  real<lower=0> sigma[J];
}

parameters {
  real mu;
  real<lower=0> tau;
  real theta[J];
}

model {
  mu ~ normal(0, 5);
  tau ~ cauchy(0, 2.5);
  theta ~ normal(mu, tau);
  y ~ normal(theta, sigma);
}
\end{verbatim}
\vspace{-7mm}
\subsubsection*{Configuration}
\vspace{-1mm}
\begin{verbatim}
./n_schools_cp sample num_warmup=100000 num_samples=1000000 
thin=1000 adapt delta=0.999 algorithm=hmc engine=nuts
max_depth=20 random seed=39457382 data file=big_schools.dat 
output file=big_schools_ehmc_cp.csv 
\end{verbatim}

\vspace{-5mm}
\subsection*{One-Way Normal (Non-Centered)}
\subsubsection*{Model}
\vspace{-5mm}
\begin{verbatim}
data {
  int<lower=0> J;
  real y[J];
  real<lower=0> sigma[J];
}

parameters {
  real mu;
  real<lower=0> tau;
  real var_theta[J];
}

transformed parameters {
  real theta[J];
  for (j in 1:J) theta[j] <- tau * var_theta[j] + mu;
}

model {
  mu ~ normal(0, 5);
  tau ~ cauchy(0, 2.5);
  var_theta ~ normal(0, 1);
  y ~ normal(theta, sigma);
}
\end{verbatim}
\vspace{-7mm}
\subsubsection*{Configuration (Baseline)}
\vspace{-1mm}
\begin{verbatim}
./n_schools_ncp sample num_warmup=5000 num_samples=100000 
adapt delta=0.99 random seed=466772400 data file=big_schools.dat 
output file=big_schools_baseline.csv 
\end{verbatim}
\vspace{-7mm}
\subsubsection*{Configuration (Nominal)}
\vspace{-1mm}
\begin{verbatim}
./n_schools_ncp sample num_warmup=5000 num_samples=50000 
adapt delta=0.8 random seed=95848382 data file=big_schools.dat 
output file=big_schools_ehmc_ncp.csv 
\end{verbatim}

\pagebreak

\bibliography{hmc_for_hier}
\bibliographystyle{PhysRevStyle}

\end{document}